\documentstyle[11pt]{article}
\begin{document}

\title{Exact solutions for the D-dimensional spherical isotropic confined harmonic oscillator}
\maketitle

\begin{center}
H. E. Montgomery Jr.,\\[0pt]
{\it Centre College,\\[0pt]
600 West Walnut Street, Danville, KY 40422 USA.}\\[0pt]
ed.montgomery@centre.edu\\[0pt]
\vglue 0.3in

G. Campoy \\[0pt]
{\it Departamento de Investigaci\'on, Universidad de Sonora,\\[0pt]
Apartado Postal A-88, 83000 Hermosillo, Sonora, M\'exico.}\\[0pt]
cajeme@cifus.uson.mx
\vglue 0.3in

N. Aquino,\\[0pt]
{\it Departamento de F\'{\i}sica, Universidad Aut\'onoma
Metropolitana--Iztapalapa,\\[0pt]
Apartado Postal 55-534, 09340 M\'exico, D.F., M\'exico}.\\[0pt]
naa@xanum.uam.mx \vglue 0.5in
\end{center}

\date{}

\begin{abstract}
We study the size effect on the energy levels of the $D$--dimensional
isotropic harmonic oscillator confined within a box of radius $r_c $ with
impenetrable walls. Two different approaches are used to obtain the energy
eigenvalues and eigenfunctions for $D$=1,2,{\ldots},5. In the first we solve
the Schr\"{o}dinger equation exactly. In the second we use a series
expansion of the wave function. The numerical results obtained are extremely
accurate; these values are reported with 50 decimal places.
\end{abstract}

\vglue 0.6in





\section{Introduction}

The idea of the spatial confinement in quantum systems has had a growing
interest in recent years, due to its potential application in the study and
production of artificial atoms in semiconductor materials and of the future
circuit devices of nano and molecular size, including the quantum computers.

\noindent On the other hand, confined quantum systems have a long history in the
modelling of a great number of applications in different areas of Physics
and Chemistry as it is shown in the cited review articles \cite{fer1}. In
the decade of the 30's, Michels, De Boer and Bijl \cite{mich} proposed the
model of a hydrogen atom confined at the centre of a sphere with
impenetrable walls and used it to study the effects of extreme pressure on
the electronic states of the hydrogen atom. This model has become one of the
most studied in the literature \cite{mich}--\cite{cecil}. The idea of
confinement inside spherical boxes has been broadly accepted and it has
continued to be used to study the electronic estates of multi-electron atoms
subject to extreme pressures \cite{zico}--\cite{cone}.

Another widely studied confined quantum system is the 1-$D$ confined
harmonic oscillator \cite{au1}--\cite{aquino3}. This system has been used as
model for the study of the proton-deuteron transformation as the energy
source in dense stars \cite{au1}--\cite{au2}, in the theory of the white
dwarfs \cite{au3} and in the escape velocity of stars from the galactic or
globular cumulus \cite{chan}. It has also been used in the study of the
specific heat of solids subjected to high pressures \cite{korson} and
magnetic properties \cite{dingle}of metals. Also few studies have been made
on the transition probabilities and Einstein coefficients for the
transitions between different levels of the 1-$D$ confined harmonic
oscillator \cite{bai}, \cite{aquino2}--\cite{aquino3}, showing that new
allowed transitions appear as a result of the confinement.

However, the harmonic oscillator confined in two and three dimensions has
received less attention \cite{aguila3}--\cite{tase2}. Recently, we have
discussed the incidental degeneracies in the 3-$D$ isotropic confined
harmonic oscillator \cite{sen}, and the conditions for the appearance of the
incidental and the inter-dimensional degeneracies in the $D$-dimensional
confined harmonic oscillator \cite{aquino5}.

In the study of the incidental and the inter-dimensional degeneracies, it is
necessary to have accurate energy eigenvalues. For the free (unconfined)
situation, the energy eigenvalues are very well known, but this is not true
for the confined problem where analytic expressions are not available and
solutions must be obtained numerically. The functional forms of the
eigenfunctions of the confined harmonic oscillators in 1, 2 and 3 dimensions
have been known for some time \cite{au1}, \cite{au2}, \cite{bai}, \cite
{aguila2}--\cite{fer2}, \cite{aguila3}. These are given in terms of the confluent
hypergeometric functions. In this work we will present two extremely precise
methods to obtain the energies of the $D$-dimensional confined oscillators.
The first is an exact method that is based on numerically obtaining the
roots of the confluent hypergeometric functions. The second method is based
on the development of the wave function in a Taylor series. This method was
used with much success previously \cite{aquino1}, \cite{aquino3}, \cite{aquino4}. 
We find that the energy eigenvalues calculated by both methods
are identical and we report them with 50 significant figures.

The content of this work is as follows: In the section 2 we present the
methods used in the obtaining of the energy. In section 3 we present our
results including comparison with previous works. Finally, in section 4 we
discuss our results and conclusions.


\section{Exact solutions}

\subsection{The one--dimensional confined HO}

The Schr\"{o}dinger equation for the one-dimensional, symmetrically confined
harmonic oscillator (in natural units where $m=\omega =\hbar =1)$ is given by

\begin{equation}  \label{eq1}
\left( {-\frac{1}{2}\frac{d^2}{dx^2}+\frac{1}{2}x^2} \right)\psi (x)=E\psi
(x),\quad
\end{equation}

\noindent where the energy is in units of $\hbar \omega $ and the unit of
the distance is $\sqrt {\hbar /m\omega } $.

The potential energy is a symmetric function of $x$, therefore the
eigenstates have definite parity; odd or even. The exact solutions are well
known \cite{au1}--\cite{au2}, \cite{bai}--\cite{hul}, \cite{dean}, \cite
{aguila2} and are obtained in terms of the Kummer or confluent
hypergeometric functions \cite{Abra}

\begin{equation}  \label{eq2}
\begin{array}{l}
\psi ^+(x)=Ae^{-x^2/2}{\ }_1F_1 \left[ {\frac{1}{4}(1-2E);\frac{1}{2};x^2} 
\right], \\ 
\psi ^-(x)=Be^{-x^2/2}x\,{\ }_1F_1 \left[ {\frac{1}{4}(3-2E);\frac{3}{2};x^2}
\right],
\end{array}
\end{equation}

\noindent where $+$ and $-$ indicate even and odd parity respectively.

In order for the wave function to be square integrable, the hypergeometric
function for the unconfined one-dimensional oscillator must terminate. This
requires that there exist some non negative integer $n$ such that

\begin{equation}  \label{eq3}
E=n+\frac{1}{2},\quad n=0,\,1,\,2,\,3,...\quad .
\end{equation}

When the harmonic oscillator is symmetrically confined in a box of length $%
2x_c $ with impenetrable walls, the energy quantization results from the
boundary conditions on the wave functions

\begin{equation}  \label{eq4}
\psi ^\pm (x=-x_c )=\psi ^\pm (x=x_c )=0.
\end{equation}

The allowed energies are obtained when the successive roots of the following
equations are found

\begin{equation}  \label{eq5}
\begin{array}{l}
{\ }_1F_1 \left[ {\frac{1}{4}\left( {1-2E} \right);\frac{1}{2};x_c^2 } 
\right]=0,\quad for\,\,even\,\,states \\ 
{\ }_1F_1 \left[ {\frac{1}{4}\left( {3-2E} \right);\frac{3}{2};x_c^2 } 
\right]=0,\quad for\,\,odd\,\,states
\end{array}
\end{equation}

To determine the energy eigenvalues, it is necessary to solve numerically
for one of the boundary conditions (\ref{eq4}) because the symmetry of the
problem. In this work we found the allowed energies using the Maple computer
algebra system and Maple's root-finding function FSOLVE. Our results are
reported in Table I.

\subsection{The D--dimensional confined HO}

The Schr\"{o}dinger equation for the isotropic harmonic oscillator in a $D$%
-dimensional Cartesian coordinate system $x_{1},x_{2}, \ldots , x_{D}$ is

\begin{equation}  \label{eq6}
\left( {-\frac{1}{2}\Delta ^{(D)}+\frac{1}{2}r^2} \right)\Psi ^{(D)}\left( {%
x_1 ,x_2 ,\ldots ,x_D } \right)=E\Psi ^{(D)}\left( {x_1 ,x_2 ,\ldots ,x_D }
\right),
\end{equation}

where $\Delta ^{(D)}$ is the $D$-dimensional Laplacian and

\begin{equation}  \label{eq7}
r^2=\sum\limits_{i=1}^D {x_i^2 } .
\end{equation}

Transforming to the $D$-dimensional spherical coordinates $\left( {r,\theta
_1 ,\theta _2 ,\ldots ,\theta _{D-1} } \right)$, we separate variables using

\begin{equation}  \label{eq8}
\Psi ^{(D)}\left( {r,\theta _1 ,\theta _2 ,\ldots ,\theta _{D-1} }
\right)=R_\ell ^{(D)} \left( r \right)\;Y_\ell ^{(D)} \left( {\theta _1
,\theta _2 \ldots ,\theta _{D-1} } \right),
\end{equation}

where $Y_\ell ^{(D)} \left( {\theta _1 ,\theta _2 \ldots ,\theta _{D-1} }
\right)$ is a normalized spherical harmonic with characteristic value $\ell
\;\left( {\ell \;+\;D\;-\;2} \right),\;\ell \;=\;0,\;1,\;2,\;\ldots $and $%
R_\ell^{\left( D \right)} \,\left( r \right)$ is a radial function that
satisfies the equation:

\begin{equation}  \label{eq9}
\left\{ {-\frac{1}{2}\left[ {\frac{d^2}{dr^2}+\frac{D-1}{r}\frac{d}{dr}-%
\frac{\ell \left( {\ell \;+\;D\;-\;2} \right)}{r^2}} \right]+\frac{1}{2}r^2}
\right\}R_\ell ^{(D)} (r)=ER_\ell ^{(D)} \left( r \right)
\end{equation}

Writing $R^{(D)}(r)=r^\ell e^{-r^2/2}F$, equation (\ref{eq9}) gives

\begin{equation}  \label{eq10}
\frac{d^2F}{dr^2}+\left( {\frac{D+2\ell -1}{r}-2r} \right)\frac{dF}{dr}%
+\left( {2E-D-2\ell } \right)F=0.
\end{equation}

Changing the variable to $z=r^2$, we obtain

\begin{equation}  \label{eq11}
z\frac{d^2F}{dz^2}+\left( {\ell +\frac{D}{2}-z} \right)\frac{dF}{dz}-\frac{1%
}{2}\left( {\ell +\frac{D}{2}-E} \right)F=0.
\end{equation}

Equation (\ref{eq11}) is the well known Kummer's differential equation \cite
{Abra}, whose regular solution at the origin is the confluent hypergeometric
function

\begin{equation}  \label{eq12}
F={\ }_1F_1 \left[ {\frac{1}{2}\left( {\ell +\frac{D}{2}-E} \right);\ell +%
\frac{D}{2};r^2} \right].
\end{equation}

In order for the wave function to be square integrable, the hypergeometric
series for the unconfined oscillator must terminate. This requirement is
satisfied if there exists some non-negative integer $n$ such that

\begin{equation}  \label{eq13}
E=2n+\ell +\frac{D}{2},\quad n=0,\,1,\,2,...\quad .
\end{equation}

When the harmonic oscillator is enclosed in an impenetrable hypersphere of
radius $r_c $, quantization results from the requirement that the radial
wave function go to zero at $r_c $. The allowed energies are found when

\begin{equation}  \label{eq14}
{\ }_1F_1 \left[ {\frac{1}{2}\left( {\ell +\frac{D}{2}-E} \right);\ell +%
\frac{D}{2};r_c^2 } \right]=0,
\end{equation}

\noindent where the successive roots are numbered $n=0,\,1,\,2,...$. As we
mentioned previously, we found the allowed energies using Maple program. Our
results for $D$=2, 3, 4 and 5 are reported in Tables II-V.


\section{Power series method}

The method that we will present has been used with much success in problems
both with and without confinement. Some of the problems solved with this
method are: the one-dimensional harmonic oscillator confined symmetrically
and asymmetrically \cite{aquino3}, the three-dimensional confined isotropic
harmonic oscillator \cite{aquino4}, the two-dimensional hydrogen atom
confined in a circle with impenetrable walls \cite{aquino6} and the
three-dimensional hydrogen atom confined in a hard sphere \cite{aquino1}.
Other applications corresponding to free problems are: the hydrogen atom
with a harmonic perturbation \cite{palma}, the quartic harmonic oscillator
and the double well potential for the inversion of NH$_3$ \cite{aquino7}, in
which the potential was represented by a polynomial of 20$^{th}$ degree.

We will describe the method briefly. For further details see references \cite
{aquino1}, \cite{aquino3}, \cite{aquino4}, \cite{aquino6}. The
Schr\"{o}dinger equation (in natural units) for one degree of freedom for an
arbitrary potential $V(x)$ can be written as

\begin{equation}  \label{eq15}
\psi ^{\prime\prime}=2\left[ {V(x)-E} \right]\quad for\;x<x_c
\end{equation}

\noindent where $x_c$ is the position of the impenetrable wall.

Now, we will suppose that the wavefunction is a function of the position $x$
and of the energy $E$.

\begin{equation}  \label{eq16}
\psi =\psi (x,E).
\end{equation}
Taking the partial derivative of equation (\ref{eq15}) respect to the energy
we obtain

\begin{equation}  \label{eq17}
{\mathop \psi \limits^\bullet}^{\prime\prime}=2\left[ {V(x)-E} \right]%
\mathop \psi \limits^\bullet -2\psi
\end{equation}

\noindent where $\mathop \psi \limits^\bullet $ denotes partial
differentiation with respect to the energy.

We need to obtain $\psi (x_c )$ and $\mathop \psi \limits^\bullet (x_c )$.
This is possible by making an initial guess $E_j $ for the value of the
energy and proceeding to integrate the equations (\ref{eq15}) and (\ref{eq17}%
). The corrected value of the energy is then obtained by means of the
Newton-Raphson formula.

\begin{equation}  \label{eq18}
E_{j+1} =E_j -\frac{\psi (x_c ,E_j )}{\mathop \psi \limits^\bullet (x_c ,E_j
)}\quad .
\end{equation}

With this new value $E_{j+1} $we calculate $\psi (x_c )$and $\mathop \psi
\limits^\bullet (x_c )$, and we use the formula (\ref{eq18}) again to obtain
a more precise value for the energy. We continue with this process until $%
\left| {E_{n+1} -E_n } \right|<\delta $, where $\delta $ is the desired
accuracy for the calculation.

The integration of the equations (\ref{eq15}) and (\ref{eq17}) is achieved
easily if we develop the wave function in a Taylor series around the origin,
where we know the initial value of the wave function $\psi (0)$.

\begin{equation}  \label{eq19}
\psi (x)=\sum_p {\frac{\psi ^{(p)}(0)}{p!}x^p} .
\end{equation}

Defining

\begin{equation}  \label{eq20}
T_p =\frac{\psi ^{(p)}(0)}{p!}x^p,
\end{equation}

then

\begin{equation}  \label{eq21}
\psi (x)=\sum\limits_p {T_p } .
\end{equation}

We can also compute $\mathop \psi \limits^\bullet (x)$ as follows

\begin{equation}  \label{eq22}
\mathop \psi \limits^\bullet (x)=\raise0.7ex\hbox{${\partial \psi }$} \!%
\mathord{\left/ {\vphantom {{\partial \psi } {\partial
E}}}\right.\kern-\nulldelimiterspace}\!\lower0.7ex\hbox{${\partial E}$}%
=\sum\limits_p {\raise0.7ex\hbox{${\partial T_p }$} \!\mathord{\left/
{\vphantom {{\partial T_p } {\partial E}}}\right.\kern-\nulldelimiterspace}\!%
\lower0.7ex\hbox{${\partial E}$}} =\sum\limits_p {\mathop {T_p }%
\limits^\bullet } .
\end{equation}

To make particular application of the method described above, we need to
calculate the coefficients $T_p$ and $\mathop{T_p }\limits^\bullet$ for each
dimension.

\subsection{The one-dimensional confined harmonic oscillator}

Substituting (\ref{eq21}) in (\ref{eq1}) the following recursion formula is
obtained for the coefficients $T_p $

\begin{equation}  \label{eq23}
T_{p+2} =\frac{2(p+1/2-E)x^2}{(p+1)(p+2)}T_p
\end{equation}

To obtain the recursion formula for the $\mathop {T_p}\limits^\bullet$
coefficients, we take the partial derivative of equation (\ref{eq21}) and
use (\ref{eq23}) to obtain

\begin{equation}  \label{eq24}
\mathop {T_{p+2}}\limits^\bullet =\frac{2[(p+1/2-E)\mathop {T_p}%
\limits^\bullet -T_p ]x^2}{(p+1)(p+2)}.
\end{equation}

For constructing the even states we used the initial conditions $\psi
(0)=1,\quad \psi ^{\prime}(0)=0$, while for the odd states the initial
conditions are $\psi (0)=0,\quad \psi ^{\prime}(0)=1$.

The other derivatives are obtained using the recurrence relationships for $%
T_p $ and $\mathop {T_p}\limits^\bullet $.\newline
The results obtained by this method are the same reported in Table I, those
are improved results of a previous work \cite{aquino3}

\subsection{ The D-dimensional confined harmonic oscillator}

Following similar steps as those described above, we found the recurrence
relations for $T_p$ and $\mathop{T_p}\limits^\bullet $:

\begin{equation}  \label{eq25}
T_{p+2} =\frac{2\,(p+l+D/2-E)r^2}{(p+2)(p+2l+D)}T_p ,
\end{equation}

and

\begin{equation}  \label{eq26}
\mathop {T_{p+2}}\limits^\bullet =\frac{2\,[(p+l+D/2-E)\mathop {T_p}%
\limits^\bullet -T_p ]r^2}{(p+2)(p+2l+D)},
\end{equation}
are obtained from Eq. (\ref{eq9}).

The results obtained by this method are reported in Tables II--V. For $D=3$
we show the improved results from reference \cite{aquino4}.


\section{Results and discussion}

\subsection{The one-dimensional confined harmonic oscillator}

As we mentioned before, the first investigators who discussed the problem of
the one-dimensional harmonic oscillator confined symmetrically in a box with
impenetrable walls were Kothari and Auluck \cite{au1}--\cite{au2}. They
found that the solutions of the Schr\"{o}dinger equation could be written in
terms of confluent hypergeometric functions. In order to obtain analytic
approaches for the energies, they used expansions and approximations in the
hypergeometric functions. In that way they obtained the correct qualitative
behaviour of the energy levels and observed that the energy values increase
quickly when diminishing the size of the box (for radius smaller than 4 au).

Baijal and Singh \cite{bai} followed a direct way and they obtained the
energies numerically by finding the zeros of an equation equivalent to (\ref
{eq5}). Their numerical results were not accurate. The reasons of this
failure are now clearly comprehensible; the absence of efficient algorithms
and computation programs to evaluate the hypergeometric functions with high
accuracy and the lack of computers that could execute these programs.
Fortunately, these impediments have now been solved in a satisfactory way.
In this work we used MAPLE, both for evaluation of the confluent
hypergeometric functions and to find the roots of the set of equations (\ref
{eq5}), following the pioneer work of Baijal and Singh \cite{bai}. We used
Maple's HYPERGEOM and FSOLVE functions and with a careful handling of the
accuracy by means of the FULLDIGITS option, we were able to obtain the
energy eigenvalues with an accuracy of 100 decimal digits.

On the other hand, by using the series method to this problem, and
programming the respective equations in the UBASIC program by using real
variables of 150 figures we calculate energy eigenvalues with 100 accuracy
figures. Our results for the ground state are reported with 50 figures in
Table I.

When comparing the energies obtained by both methods described above to 100
figures we found that the relative error between both calculations are
smaller than $1\times 10^{-100}$. This shows that the two quite different
methods are both very stable and accurate. To our knowledge, these are the
most precise calculations that have been reported until now.

\subsection{The {\em 2--D} and {\em 3--D} confined harmonic oscillators}

The exact formulation of the problem of the 2-$D$ and 3-$D$ confined
harmonic oscillator was proposed by Aguilera-Navarro et al \cite{aguila3} in
1983. They obtained a transcendental equation in terms of confluent
hypergeometric functions, whose roots are the energy eigenvalues, but they
didn't solve it, noting that the numeric solution of that equation required
hard computational effort. As in the one-dimensional problem, the
researchers decided to use other approaches and methods such as the
following: i) perturbation theory, ii) Pad\'{e} approximants and iii) direct
diagonalization of the Hamiltonan matrix in the basis set of the free
particle, finding the matrix elements analytically. The results were
obtained with 6 decimal places by diagonalizing matrixes up to 50x50. Those
results were at that time considered to be the exact result.

Taseli and Zafer \cite{tase1} used essentially the same method that
Aguilera-Navarro et al, for studying the harmonic oscillator and polynomial
potentials. Taseli and Zafer report their results with 30 decimal places.
For comparison, their results are shown in Table II for the 2-$D$ harmonic
oscillator and in Table III for 3-$D$ problem.

Our results obtained by solving equation (\ref{eq14}) numerically using
MAPLE and using the series method are reported with 50 decimal places for
the ground state of the 2-$D$ and 3-$D$ confined harmonic oscillators in
Table II and Table III respectively. The results obtained by these two
methods coincided up to 100 significant figures.

\subsection{The {\em 4--D} and {\em 5--D} confined harmonic oscillators.}

Our results obtained by numerically solving the set of equations (\ref{eq14}%
) by means of MAPLE program and by using the series method are reported with
50 significant figures, for the ground state of the 4-$D$ and 5-$D$ confined
harmonic oscillators in Table IV and Table V respectively. The results
obtained by these two methods coincide up to 100 decimal places. This is the
first time that energy eigenvalues for the 4-$D$ and 5-$D$ confined harmonic
oscillators have been reported.

\section{Final remarks}

Kotari and Auluck \cite{au1}--\cite{au2} formulated very early the problem
in a correct way. Baijal and Singh \cite{bai} solved the problem
numerically; however the accuracy in their results was limited to 1 part in
10,000. Due to the difficulties of obtaining the solution of this problem
through the use of the hypergeometric functions, later researchers opted to
use different methods and approaches. In this work we have shown that by
using the hypergeometric functions of MAPLE computer algebra system, it is
possible to obtain very accurate results. The accuracy of this method was
confirmed by the series method described above. The coincidence of the
results of both methods up to 100 decimal places shows that both approaches
are very stable and accurate. The present results provide the benchmark to
probe the accuracy of new methods.

\vglue 0.5in

\noindent {\bf Acknowledgments}

\noindent We thank K. D. Sen for his comments on this work

\pagebreak


\pagebreak
\renewcommand{\baselinestretch}{0.96}
 
\begin{table}
Table I. Ground state and first excited energies of the one-dimensional
harmonic oscillator as a function of the box length $2x_c$. The energies are
in units of $\hbar \protect\omega$ and the distances in units of $\protect%
\sqrt{\hbar/m \protect\omega}$. \\

\begin{tabular}{ll}
\hline
$x_{c}$ & Ground state energy \\ \hline
0.5 & 4.9511293232 5413041195 1134080515 9857388997 9644551237 \\ 
1.0 & 1.2984598320 3205669378 4230206450 2370269582 3439869300 \\ 
1.5 & 0.6889317536 4684808297 9408577127 0594543020 2986219232 \\ 
2.0 & 0.5374612092 8167516049 2498062617 3690881483 5739573494 \\ 
2.5 & 0.5049541046 2368718050 9447174762 0176047282 7339258667 \\ 
3.0 & 0.5003910829 2974859059 4328365218 4059730111 1218415624 \\ 
3.5 & 0.5000180448 2030945381 6616486788 1801095234 3877637574 \\ 
4.0 & 0.5000004908 5643052761 7123693880 1651645285 4015551420 \\ 
4.5 & 0.5000000079 3815418303 1804233314 0697872385 1651045940 \\ 
5.0 & 0.5000000000 7671713198 9118613783 6722860290 6861259474 \\ 
6.0 & 0.5000000000 0000154791 6958282084 2671674954 5402080427 \\ 
7.0 & 0.5000000000 0000000000 4098007117 5362147129 8519793926 \\ 
8.0 & 0.5000000000 0000000000 0000001436 2707054755 7659037566 \\ 
9.0 & 0.5000000000 0000000000 0000000000 0000670071 2965313260 \\ 
10.0 & 0.5000000000 0000000000 0000000000 0000000000 0041764526 \\ 
11.0 & 0.5000000000 0000000000 0000000000 0000000000 0000000000 \\ \hline
$x_{c}$ & First excited state energy \\ \hline
0.5 & 19.7745341792 0831989835 4604517172 0308265073 6578782343 \\ 
1.0 & 5.0755820152 2678306601 7648991449 8809070971 4847782792 \\ 
1.5 & 2.5049761785 3502402045 9213743876 6702242013 4355616840 \\ 
2.0 & 1.7648164387 8063679020 2259586613 1246579254 5753601492 \\ 
2.5 & 1.5514216545 5474477980 5797055037 4466896712 3648964272 \\ 
3.0 & 1.5060815272 5279462165 2764212526 2481287731 9700177501 \\ 
3.5 & 1.5003995211 9607101202 1916274934 7869410041 1346943413 \\ 
4.0 & 1.5000146030 0712398734 9091703023 5066898625 6879252750 \\ 
4.5 & 1.5000003041 6594363224 6948818889 9660548783 0443971120 \\ 
5.0 & 1.5000000036 7158393112 6083763054 5056423792 2132790701 \\ 
6.0 & 1.5000000000 0010821056 8920426294 9699221373 4008991988 \\ 
7.0 & 1.5000000000 0000000039 3137796646 4980630573 4262919050 \\ 
8.0 & 1.5000000000 0000000000 0000180898 7877745393 8765502196 \\ 
9.0 & 1.5000000000 0000000000 0000000000 0107185446 9753799582 \\ 
10.0 & 1.5000000000 0000000000 0000000000 0000000000 8268077088 \\ 
11.0 & 1.5000000000 0000000000 0000000000 0000000000 0000000008 \\ \hline
\end{tabular}
\end{table}
\pagebreak
\begin{table}
Table II. Energy eigenvalues for the two-dimensional isotropic confined
harmonic oscillator for $n=0, l=0$ and $n=0, l=1$ as a function of the
confinement radius $r_c$ , and its comparison with Taseli's results$^a$
Ref.[41]. The energies are in units of $\hbar \omega$ and the distances in
units of $\sqrt{\hbar/m \omega}$.\\

\begin{tabular}{ll}
\hline
$r_c$ & $n=0, l=0$ \\ \hline
0.5 & 11.5936192506 8668479643 2170521897 6492016024 7818580170 \\ 
1.0 & 3.0000000000 0000000000 0000000000 0000000000 0000000000 \\ 
1.5 & 1.5235322602 5914873036 6271226432 8839654005 8580817118 \\ 
2.0 & 1.1222085296 7891837492 4623606583 7385347311 4764534647 \\ 
2.5 & 1.0199306851 0149764966 9029200736 9635571918 5241114989 \\ 
3.0 & 1.0019367879 6432851707 9965303704 9698536351 6685928166 \\ 
3.5 & 1.0001065838 7526319243 0778389006 5275059562 5162790712 \\ 
4.0 & 1.0000033582 1521855746 5783438694 2835633641 8541234989 \\ 
4.5 & 1.0000000616 1214749775 2604428805 1580090716 1945112286 \\ 
5.0 & 1.0000000006 6534686266 8798296163 8488876290 9044905310 \\ 
5.0$^a$ & 1.0000000006 65385 \\ 
6.0 & 1.0000000000 0001622254 6342645858 8254746724 7650899752 \\ 
7.0 & 1.0000000000 0000000005 0308861850 7863587003 2026301379 \\ 
7.0$^a$ & 1.0000000000 0000000005 03 \\ 
8.0 & 1.0000000000 0000000000 0000020202 7529579627 9788625777 \\ 
9.0 & 1.0000000000 0000000000 0000000000 0010621779 8865980561 \\ 
9.0$^a$ & 1.0000000000 0000000000 0000000000 00 \\ 
10.0 & 1.0000000000 0000000000 0000000000 0000000000 0736498307 \\ 
11.0 & 1.0000000000 0000000000 0000000000 0000000000 0000000001 \\ 
11.0$^a$ & 1.0000000000 0000000000 0000000000 00 \\ \hline
$r_c$ & $n=0, l=1$ \\ \hline
0.5 & 29.4056004466 9756111909 4989312607 2533606944 3885645082 \\ 
1.0 & 7.5071721804 5194296125 5970569322 4683728514 7610899602 \\ 
1.5 & 3.6322191884 0214310001 3125848888 3649489632 2914146137 \\ 
2.0 & 2.4717752113 5017965087 7413015393 1937987026 1062386895 \\ 
2.5 & 2.1057033473 6386137979 9168786765 6763723197 9306909584 \\ 
3.0 & 2.0149671135 0308735580 2990773059 6826969574 0423889838 \\ 
3.5 & 2.0011722745 5370550365 2986053259 5949083728 5949607461 \\ 
4.0 & 2.0000497838 2870035813 8418130090 9885102064 2824637025 \\ 
4.5 & 2.0000011782 3371178944 8218912905 1701308538 1618885601 \\ 
5.0 & 2.0000000159 0394948216 9465718866 9200664238 5393456486 \\ 
6.0 & 2.0000000000 0056676874 0232254818 0017888510 4212103477 \\ 
7.0 & 2.0000000000 0000000241 2582416245 4898871780 6700297398 \\ 
8.0 & 2.0000000000 0000000000 0001272098 8438202771 4243792590 \\ 
9.0 & 2.0000000000 0000000000 0000000000 0849466212 9702927122 \\ 
9.0$^a$ & 2.0000000000 0000000000 000000000 \\ 
10.0 & 2.0000000000 0000000000 0000000000 0000000007 2897978486 \\ 
11.0 & 2.0000000000 0000000000 0000000000 0000000000 0000000081 \\ \hline
\end{tabular}
\end{table}
\pagebreak
\begin{table}
Table III. Energy eigenvalues for the three-dimensional isotropic confined
harmonic oscillator for $n=0, l=0$ and $n=0, l=1$ as a function of the
confinement radius $r_c$, and its comparison with Taseli's results$^b$
Ref.[43]. The energies are in units of $\hbar \protect\omega$ and the
distances in units of $\protect\sqrt{\hbar/m \protect\omega}$.\\

\begin{tabular}{ll}
\hline
$r_c$ & $n=0, l=0$ \\ \hline
0.5 & 19.7745341792 0831989835 4604517172 0308265073 6578782343 \\ 
1.0 & 5.0755820152 2678306601 7648991449 8809070971 4847782792 \\ 
1.5 & 2.5049761785 3502402045 9213743876 6702242013 4355616840 \\ 
2.0 & 1.7648164387 8063679020 2259586613 1246579254 5753601492 \\ 
2.5 & 1.5514216545 5474477980 5797055037 4466896712 3648964272 \\ 
3.0 & 1.5060815272 5279462165 2764212526 2481287731 9700177501 \\ 
3.5 & 1.5003995211 9607101202 1916274934 7869410041 1346943413 \\ 
4.0 & 1.5000146030 0712398734 9091703023 5066898625 6879252750 \\ 
4.5 & 1.5000003041 6594363224 6948818889 9660548783 0443971120 \\ 
5.0 & 1.5000000036 7158393112 6083763054 5056423792 2132790701 \\ 
5.0$^b$ & 1.5000000036 715 \\ 
6.0 & 1.5000000000 0010821056 8920426294 9699221373 4008991988 \\ 
7.0 & 1.5000000000 0000000039 3137796646 4980630573 4262919050 \\ 
7.0$^b$ & 1.5000000000 0000000039 315 \\ 
8.0 & 1.5000000000 0000000000 0000180898 7877745393 8765502196 \\ 
9.0 & 1.5000000000 0000000000 0000000000 0107185446 9753799582 \\ 
9.0$^b$ & 1.5000000000 0000000000 000000000 \\ 
10.0 & 1.5000000000 0000000000 0000000000 0000000000 8268077088 \\ 
11.0 & 1.5000000000 0000000000 0000000000 0000000000 0000000008 \\ 
11.0$^b$ & 1.5000000000 0000000000 000000000 \\ \hline
$r_c$ & $n=0, l=1$ \\ \hline
0.5 & 40.4282764968 8303569286 7657198337 4600503452 5299138678 \\ 
1.0 & 10.2822569391 5401409565 3163129127 6677765073 6003398500 \\ 
1.5 & 4.9035904194 0884107768 8890567799 6737146733 1869143993 \\ 
2.0 & 3.2469470987 7100992231 2836277400 7282559791 2938650857 \\ 
2.5 & 2.6881439638 9023726499 4772298426 8173993948 1291108167 \\ 
3.0 & 2.5312924666 1555916863 7233915188 7007997989 8790276246 \\ 
3.5 & 2.5029101642 9565984306 7531899942 2032477828 3713066610 \\ 
4.0 & 2.5001437781 6983615678 2542969854 8358197775 8222780459 \\ 
4.5 & 2.5000038701 0746391479 0391925903 5488977270 8091556844 \\ 
5.0 & 2.5000000584 4093459377 8654916228 3700691294 3132687962 \\ 
6.0 & 2.5000000000 0251914362 4008272991 8386092490 2462496073 \\ 
7.0 & 2.5000000000 0000001256 5630089465 7702336239 3748249742 \\ 
8.0 & 2.5000000000 0000000000 0007592671 4601199360 0262516226 \\ 
9.0 & 2.5000000000 0000000000 0000000000 5714219098 1734768092 \\ 
10.0 & 2.5000000000 0000000000 0000000000 0000000054 5548679893 \\ 
11.0 & 2.5000000000 0000000000 0000000000 0000000000 0000000669 \\ \hline
\end{tabular}
\end{table}
\pagebreak
\begin{table}
Table IV. Energy eigenvalues for the four-dimensional isotropic confined
harmonic oscillator for $n=0, l=0$ and $n=0, l=1$ as a function of the
confinement radius $r_c$ . The energies are in units of $\hbar \protect\omega
$ and the distances in units of $\protect\sqrt{\hbar/m \protect\omega}$.\\

\begin{tabular}{ll}
\hline
$r_c$ & $n=0, l=0$ \\ \hline
0.5 & 29.4056004466 9756111909 4989312607 2533606944 3885645082 \\ 
1.0 & 7.5071721804 5194296125 5970569322 4683728514 7610899602 \\ 
1.5 & 3.6322191884 0214310001 3125848888 3649489632 2914146137 \\ 
2.0 & 2.4717752113 5017965087 7413015393 1937987026 1062386895 \\ 
2.5 & 2.1057033473 6386137979 9168786765 6763723197 9306909584 \\ 
3.0 & 2.0149671135 0308735580 2990773059 6826969574 0423889838 \\ 
3.5 & 2.0011722745 5370550365 2986053259 5949083728 5949607462 \\ 
4.0 & 2.0000497838 2870035813 8418130090 9885102064 2824637025 \\ 
4.5 & 2.0000011782 3371178944 8218912905 1701308538 1618885601 \\ 
5.0 & 2.0000000159 0394948216 9465718866 9200664238 5393456486 \\ 
6.0 & 2.0000000000 0056676874 0232254818 0017888510 4212103477 \\ 
7.0 & 2.0000000000 0000000241 2582416245 4898871780 6700297398 \\ 
8.0 & 2.0000000000 0000000000 0001272098 8438202771 4243792590 \\ 
9.0 & 2.0000000000 0000000000 0000000000 0849466212 9702927122 \\ 
10.0 & 2.0000000000 0000000000 0000000000 0000000007 2897978486 \\ 
11.0 & 2.0000000000 0000000000 0000000000 0000000000 0000000081 \\ \hline
$r_c$ & $n=0, l=1$ \\ \hline
0.5 & 52.8003728316 3788346093 0950305785 0051346372 5631801073 \\ 
1.0 & 13.3915380494 6480185532 0494009283 4067653436 0714152088 \\ 
1.5 & 6.3173223844 6528516940 0939563655 0086913973 3614141779 \\ 
2.0 & 4.0925993533 4910526414 2040215854 0583516233 3845977779 \\ 
2.5 & 3.3030049737 7100747132 4786533205 4015439195 2504852928 \\ 
3.0 & 3.0580504736 2975666805 2943138274 5161600133 0555173037 \\ 
3.5 & 3.0063652100 0233518767 0490718817 2904327683 1566946907 \\ 
4.0 & 3.0003661001 4285621959 4091284711 1602469471 7391623416 \\ 
4.5 & 3.0000112193 8847681098 4494750875 7522162172 5417870407 \\ 
5.0 & 3.0000001896 2579523225 9655764522 1335258633 9265148349 \\ 
6.0 & 3.0000000000 0989077737 6549090900 3787454466 4191972235 \\ 
7.0 & 3.0000000000 0000005781 9369976287 6503055609 2997874251 \\ 
8.0 & 3.0000000000 0000000000 0040038694 5911963772 5227747959 \\ 
9.0 & 3.0000000000 0000000000 0000000003 3961858787 5639028352 \\ 
10.0 & 3.0000000000 0000000000 0000000000 0000000360 7297869310 \\ 
11.0 & 3.0000000000 0000000000 0000000000 0000000000 0000004872 \\ \hline
\end{tabular}
\end{table}
\pagebreak
\begin{table}
Table V. Energy eigenvalues for the five-dimensional isotropic confined
harmonic oscillator for $n=0, l=0$ and $n=0, l=1$ as a function of the
confinement radius $r_c$ . The energies are in units of $\hbar \protect\omega
$ and the distances in units of $\protect\sqrt{\hbar/m \protect\omega}$.\\

\begin{tabular}{ll}
\hline
$r_c$ & $n=0, l=0$ \\ \hline
0.5 & 40.4282764968 8303569286 7657198337 4600503452 5299138678 \\ 
1.0 & 10.2822569391 5401409565 3163129127 6677765073 6003398500 \\ 
1.5 & 4.9035904194 0884107768 8890567799 6737146733 1869143993 \\ 
2.0 & 3.2469470987 7100992231 2836277400 7282559791 2938650857 \\ 
2.5 & 2.6881439638 9023726499 4772298426 8173993948 1291108167 \\ 
3.0 & 2.5312924666 1555916863 7233915188 7007997989 8790276246 \\ 
3.5 & 2.5029101642 9565984306 7531899942 2032477828 3713066610 \\ 
4.0 & 2.5001437781 6983615678 2542969854 8358197775 8222780459 \\ 
4.5 & 2.5000038701 0746391479 0391925903 5488977270 8091556844 \\ 
5.0 & 2.5000000584 4093459377 8654916228 3700691294 3132687963 \\ 
6.0 & 2.5000000000 0251914362 4008272991 8386092490 2462496073 \\ 
7.0 & 2.5000000000 0000001256 5630089465 7702336239 3748249742 \\ 
8.0 & 2.5000000000 0000000000 0007592671 4601199360 0262516226 \\ 
9.0 & 2.5000000000 0000000000 0000000000 5714219098 1734768092 \\ 
10.0 & 2.5000000000 0000000000 0000000000 0000000054 5548679893 \\ 
11.0 & 2.5000000000 0000000000 0000000000 0000000000 0000000669 \\ 
$r_c$ & $n=0, l=1$ \\ \hline
0.5 & 66.4897565362 4517756305 3308770820 2330012102 2691586465 \\ 
1.0 & 16.8277771096 2480257625 9617469161 0755246538 5604625452 \\ 
1.5 & 7.8717304877 6674618813 0354341233 8621971714 2326103755 \\ 
2.0 & 5.0100408656 3599642760 4228631968 4014661533 4471780425 \\ 
2.5 & 3.9535289034 8568636015 5314458749 0430486562 2699519238 \\ 
3.0 & 3.5982476989 7005161570 7481560771 7287260251 5760278006 \\ 
3.5 & 3.5125803181 1996483612 8255886902 6750312869 7349417658 \\ 
4.0 & 3.5008420738 2978240548 0678521299 6397537393 6306856269 \\ 
4.5 & 3.5000294123 9950862348 5874703325 8737188786 0817955259 \\ 
5.0 & 3.5000005567 1516742638 0149068727 2109923147 0235826666 \\ 
6.0 & 3.5000000000 3515113957 7921310576 3564828467 8442048896 \\ 
7.0 & 3.5000000000 0000024085 3639264861 3384517091 1914799496 \\ 
8.0 & 3.5000000000 0000000000 0191153117 0090662417 7345620967 \\ 
9.0 & 3.5000000000 0000000000 0000000018 2748879438 1689931231 \\ 
10.0 & 3.5000000000 0000000000 0000000000 0000002159 5637174550 \\ 
11.0 & 3.5000000000 0000000000 0000000000 0000000000 0000032115 \\ \hline
\end{tabular}
\end{table}

\end{document}